\documentstyle[floats,epsfig,aps]{revtex}

\newcommand {\be} {\begin{equation}} 
\newcommand {\ee} {\end{equation}} 
\newcommand {\Be}{\begin{eqnarray*}}
\newcommand {\Ee} {\end{eqnarray*}}
\newcommand {\bey} {\begin{eqnarray}} 
\newcommand {\eey} {\end{eqnarray}} 
\newcommand{\za}{\alpha}

\newcommand{\zl}{\lambda}

\newcommand{\zt}{\tau}

\newcommand{\zN}{I\hskip-3.4pt N}

\newcommand{\veps}{\varepsilon}

\begin{document}
\begin{center}
\LARGE{\bf The Gallavotti-Cohen Fluctuation Theorem \\
for a non-chaotic model}\\
\end{center}
\vspace{0.3cm}
\begin{center}
\large{
S.\ Lepri~$^1$
\footnote{also at INFM - Unit\`a di Firenze}
,~~ L. Rondoni~$^2,~~$G.\ Benettin~$^3$
\footnote{also at GNFM-CNR and INFM - Unit\`a di Padova}
} \end{center}
\begin{center}
\small{\it 1. Dip. Energetica ``S. Stecco'', Universit\'a di Firenze
Via S. Marta 3 I-50139 Firenze, Italy \\
e-mail: {\rm lepri@avanzi.de.unifi.it} \\
2. Dip. Matematica, Politecnico di Torino
Corso Duca degli Abruzzi 24, I-10129 Torino, Italy \\ 
e-mail:  {\rm rondoni@polito.it} \\
3. Dip. Matematica Pura e Applicata, Universit\'a di Padova
via Belzoni 7 I-35131, Padova, Italy\\
}
\end{center}

\vspace{0.5cm}
\centerline{\small (\today)}

\vspace{1cm}

\section*{Abstract}
{\small We test the applicability of the Gallavotti-Cohen fluctuation 
formula on a nonequilibrium version of the periodic Ehrenfest wind-tree 
model. This is a one-particle system whose dynamics is rather complex 
(e.g.\ it appears to be diffusive at equilibrium), but its Lyapunov 
exponents are nonpositive. 
For small applied field, the system exhibits a very long transient,
during which the dynamics is roughly chaotic, followed by asymptotic collapse
on a periodic orbit. During the transient, the dynamics is diffusive, 
and the fluctuations of the current 
are found to be in agreement with the fluctuation formula, despite the  
lack of real hyperbolicity. These results also constitute 
an example which manifests the difference between the 
fluctuation formula and the Evans-Searles identity.}

\section{Introduction}
In molecular dynamics simulations of fluids in nonequilibrium 
stationary states, Evans, Cohen and Morriss \cite{ECM2} discovered 
a remarkable relation for the fluctuations of the entropy production
rate. This relation links in a striking fashion the microscopic 
reversible dynamics of certain particle systems in a stationary state, 
to the corresponding irreversible macroscopic dynamics. Inspired by
these findings, Gallavotti and Cohen proved \cite{GC} the 
fluctuation relation for a wide class of systems, directly from 
the dynamics of their constituent particles. Their proof was based 
on the following:

\vskip 5pt \noindent 
{\it {\bf Chaotic Hypothesis (CH):} A reversible many-particle 
system in a stationary state can be regarded 
as a transitive Anosov system for the purpose of computing
its macroscopic properties.} 

\vskip 5pt \noindent
The ensuing result is now known as the Gallavotti-Cohen
Fluctuation Theorem (GCFT). Near equilibrium, the GCFT implies both the 
Onsager and Einstein relations \cite{GGOR} and can therefore be interpreted 
as an extension of them to far-from-equilibrium situations. 

On the other hand, quoting from the review article \cite{GG3}: 
{\em ``... in concrete cases not only it is not known whether the system 
is Anosov but, in fact, it is usually clear that it is not ... Hence the 
test is necessary to check the CH which says that the failure of the 
Anosov property should be irrelevant for practical purposes.''} 
For ``practical purposes'' means that the calculation of quantities of 
physical interest is not affected by the deviations of the dynamics from the 
ideal case of an Anosov flow. Therefore, numerical or real experiments are 
required to test the applicability of the CH, and to identify its range 
of validity. Several papers have been devoted to this purpose (see, e.g.\ 
Refs.\cite{BGG,CL,LLP,BCL,RS}), while other papers have investigated the 
possibility of observing fluctuation relations similar to that of the GCFT 
in different contexts (see e.g.\ \cite{GG1,BG,CM,LS}). 
These works show that the CH is appropriate in the interpretation of: 
(a) the numerical results obtained for two-dimensional systems of hard-core 
particles \cite{BGG,BCL}; (b) some experiment on liquids undergoing 
Benard convection \cite{CL}; (c) numerical simulations of two-dimensional 
turbulent fluids \cite{RS}; (d) the heat transport along chains of 
anharmonic oscillators \cite{LLP}. At the same time, the 
papers \cite{CM,LS} extend the validity of the GCFT to stochastic dynamics, 
including rather general Markov processes. 

>From all the mentioned examples, one can indeed conclude that --according to
the original intuition of Ref.\cite{GC} -- the CH effectively works for a 
definitely wider class of systems than that of topologically 
mixing Anosov diffeomorphisms or flows. Of course, despite of the lack of
uniform hyperbolicity (or smoothness, or both) all these time-reversal 
invariant systems share the common property of being strongly chaotic, in 
the sense that they have (possibly many) positive Lyapunov exponents.
\footnote{Strictly speaking, such remark applies only to deterministic
dynamics: nonetheless, stochastic systems are consistent with the presence 
of chaotic underlying dynamics}

One question comes to the fore: given a time-reversible dynamical 
system, what is the minimal degree of ``complexity'' required for its microscopic
dynamics to verify the fluctuation relation? Alternatively, one may ask
how ``weakly chaotic'' can be a system which verifies the CH.

These are rather natural questions in the framework of statistical
mechanics, where similar problems have traditionally been considered.
For instance, the assumptions of ergodicity or of molecular chaos,  
are universally accepted for equilibrium systems, in spite of the well 
known fact that such properties are not only exceedingly difficult to 
prove in practical cases, but are violated in most of them.
Nevertheless, these assumptions lead to the correct physical predictions, 
and provide a mechanical foundation to thermodynamics by linking the latter
to the microscopic dynamics. The reasons of their success 
are hidden in the interplay of extremely different time and length 
scales, and in the large numbers of particles which constitute macroscopic
systems (see, e.g., \cite{GG3,JLL,Br,CR} for a discussion of these 
topics, and Ref.\cite{BHS} for a recent work on the role of
different time scales in classical gases). 

In the present paper we approach those issues by studying a modified 
version of the Ehrenfest wind-tree model. It consists of a particle 
bouncing elastically in an array of fixed polygonal scatterers.
The original version (randomly distributed and square obstacles) was
introduced to study the validity of Boltzmann equation and has been
very recently reconsidered \cite{CPD}.  
The most remarkable properties for our purposes is the fact that flat 
boundaries prevent chaotic behaviour as no defocussing of nearby 
trajectories occours. Nonetheless, the model has ``good'' statistical
properties: the $H$-theorem holds, and the moving particle 
gives rise to a true brownian motion.

What will be considered here is a nonequilibrium version of the model,
with an external field and a Gaussian thermostat (see, for the details,
the next section). The behavior of the system for nonvanishing field is
essentially the following: asymptotically the dynamics is trivial,
namely the particle collapses onto a periodic orbit. Before this,
however, there is a long transient (of duration
growing as $\veps^{-2}$, if $\veps$
denotes the intensity of the applied field) during which the motion is
basically diffusive as in the equilibrium case. Our main result is that
{\it the} GCFT {\it holds during such a transient}. One may therefore
speculate that, although chaoticity is in principle necessary to
guarantee the validity of the GCFT for all times, the latter may still
retain its meaning even in the absence of chaoticity, for trajectories 
of large although finite length.

Beyond this, our results also contribute, in our opinion, to clarify questions
which have been recently raised in the literature. First of all, they provide
an example in which the different origin of the GCFT and of a previous 
result, known as the Evans-Searles Identity \cite{ESI}, are evident 
(cf.\ Section III). Moreover, our findings give further support to the claim
\cite{CPD} that observing diffusive behaviour does not constitute
by itself a proof of the chaoticity of the microscopic dynamics \cite{GBetc}
(cf.\ Section II).

\section{The modified Ehrenfest gas}

\begin{figure}
\centering\epsfig{figure=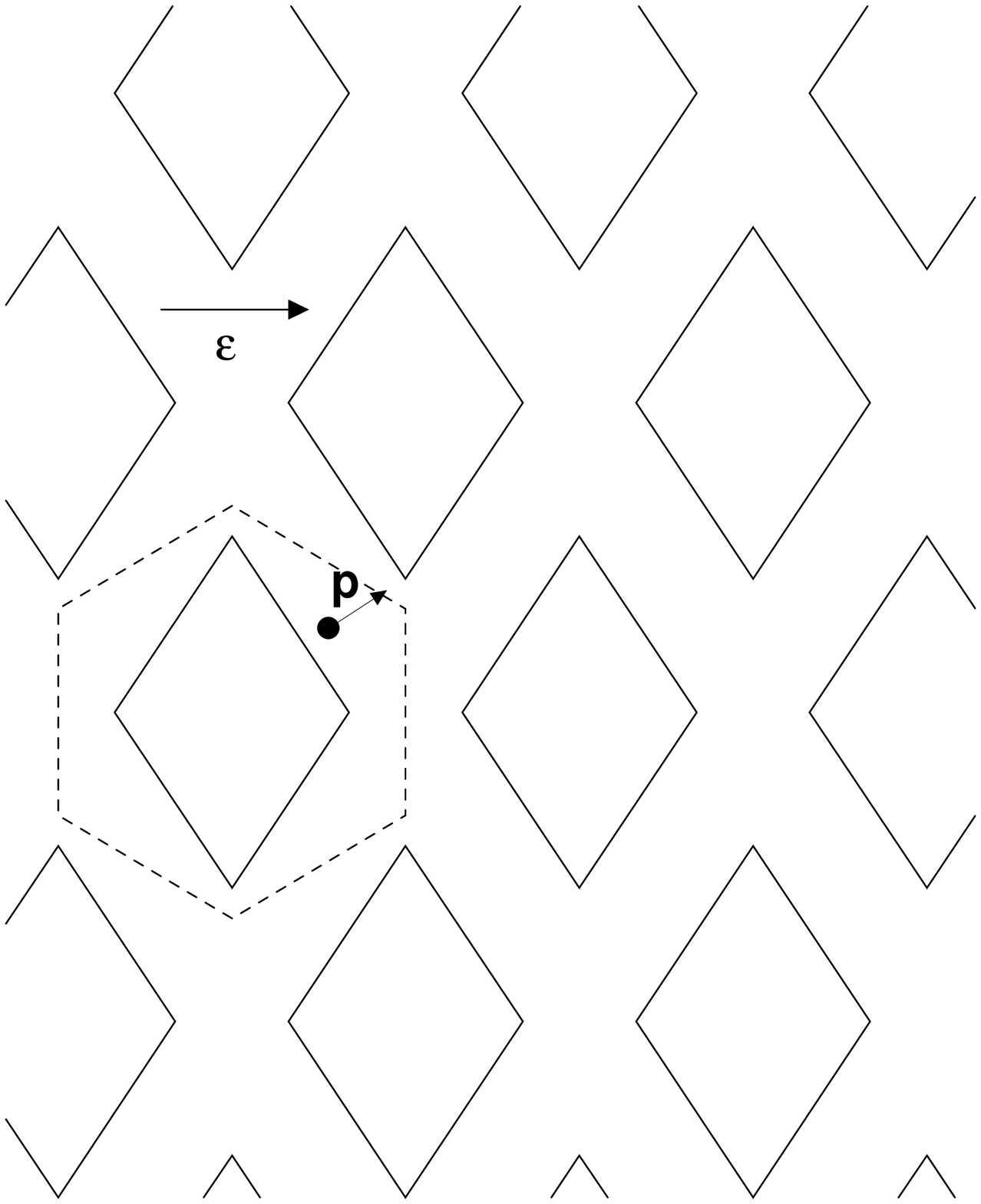,width=9cm}
\vskip 1cm
\caption{\baselineskip=12pt \small
The modified Ehrenfest gas. Along the paper, the side of the elementary
cell is set to 1.291, while the semiaxis of the rhombus are chosen to be 
1.1 and 0.7573 respectively.}
\label{lattice}
\end{figure}

The periodic Ehrenfest gas consists of one moving particle 
of mass $m$, which is elastically scattered by a set of rhomboidal
fixed obstacles arranged on a  triangular lattice (see fig.\ \ref{lattice}). 
Between any two collisions, the particle  moves freely along straight 
lines. The periodic structure of the lattice allows us to follow 
the motion of the particle, looking at the periodic image of its trajectory 
in just one hexagonal cell: the {\em elementary cell} evidenced in fig.\
\ref{lattice}. 
As the sides of the scatterers are flat, the dynamics cannot be chaotic: 
all of the the four Lyapunov exponents $\lambda_i$ vanish. Nevertheless, for
generic (i.e.\ irrational) values of the internal angles of the rhombus, 
the system is expected to be ergodic, so that in particular a
long enough trajectory fills the available phase space 
(see \cite{gutkin} for a review on the subject, and \cite{artuso}
for more recent results). 

Similarly to the case of the nonequilibrium Lorentz gas 
\cite{lamberto}, we modify this model by adding an external field of 
intensity $\varepsilon$, and introduce a Gaussian thermostat, 
which constraints the kinetic 
energy of the particle to its initial value $K$. Let $(x,y)$ be the position 
of the particle, and $(p_x,p_y)$ be its momentum. We fix $m=1$, $K=1/2$ and 
let the field  point in the positive $x$-direction. The equations of motion 
for the free flights thus read
\be
\left\{ \begin{array}{ll}
\dot x = p_x ~; & \quad \dot p_x =  - \alpha p_x + \varepsilon\\
\dot y = p_y ~; & \quad \dot p_y =  - \alpha p_y \end{array}
\right. \quad \mbox{with} \quad \za = \veps p_x~.
\label{motion}
\ee
The effect of the external force is that the segments of trajectory between 
subsequent collisions get curved, but because of the simple form of 
Eqs.(\ref{motion}) 
they can be computed analytically (see the Appendix).

We performed numerical simulations of the model 
for small and moderate fields, $10^{-4} < \veps < 1$ by evolving 
eqs.\ (\ref{motion}) starting from generic initial conditions. The asymptotic 
state  was always found to be a periodic orbit.\footnote{We recall that 
periodic orbits in the elementary cell may be open in full phase space, 
with a total displacement of an integer number of lattice vectors for each 
period.} 
Accordingly, upon switching on the field, two of the four Lyapunov exponents 
remain zero (the exponents corresponding respectively to the 
conserved kinetic energy,
$\zl_1$, and to the direction of the flow, $\zl_2$) while the other
two are numerically seen to approach a negative or vanishing value: 
\be
0 = \lambda_1 \; = \; \lambda_2 \; \ge \; \lambda_3 > \lambda_4 \quad.
\label{estim}
\ee
In practice, for small fields like those mostly considered here, the exponent
$\lambda_3$ is often so small that numerically we can 
hardly distinguish it from zero, while $\zl_4$ takes definitely 
negative values.

The mere existence of a trivial asymptotic motion for $\veps\ne0$  
does not however exclude quite complicated behaviour. Indeed, as dissipation 
is weak for small $\varepsilon$, a long transient is required to reach the 
periodic orbit. During the transient the motion of the particle looks rather 
erratic, and covers almost uniformly a large fraction of the phase space 
$\Omega$. Furthermore, the behaviour of the system on this time scale appears 
to be almost stationary, and can be described in a statistical way. 

To illustrate these facts, and  visualize the dynamics, it is convenient to 
introduce the usual ``bounce map'' of billiards. Precisely, each collision is 
assigned coordinates $(s,\cos\psi)$, where $s$ is the distance of the 
collision point from, say, the rightmost angle of the scatterer, 
measured along its perimeter, and $\psi$ is the angle between the outcoming 
momentum and the side of the rhombus (positively oriented in the 
counterclockwise direction); the billiard dynamics then defines a map 
\be
B: \quad (s_n,\cos\psi_n) ~\quad \mapsto \quad (s_{n+1},\cos\psi_{n+1})\ ,
\ee
where $(s_n,\cos\psi_n)$ denote the coordinates of the $n$--th 
collision. For $\varepsilon = 0$,
the map turns out to be area preserving, so that ergodicity 
corresponds to uniform filling of the square (or, better, of the cylinder) 
$[0,L)\times[-1,1]$, $L$ denoting the overall lenghth of the border. 

\begin{figure}
\centering\epsfig{figure=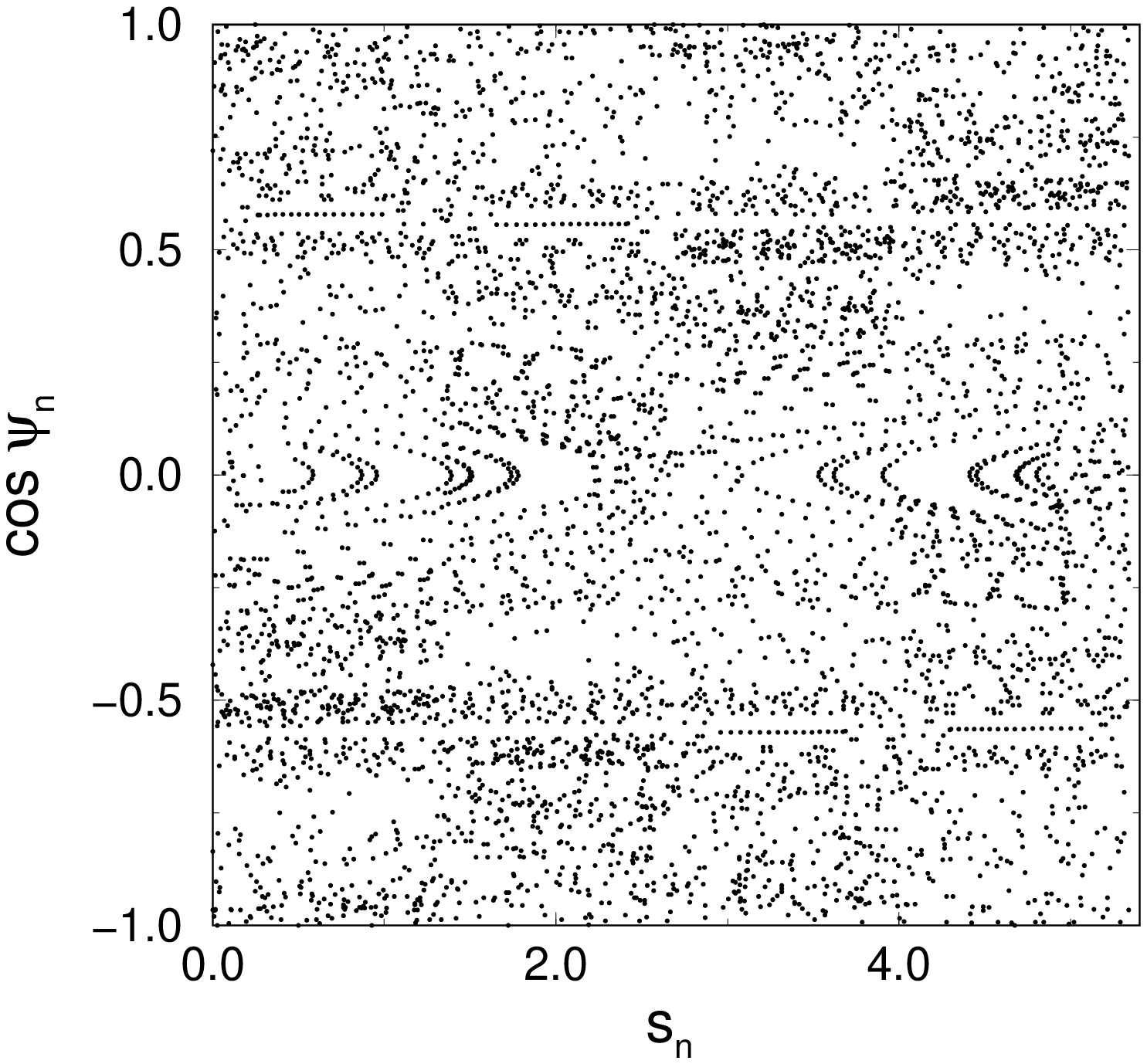,width=7cm}
\centering\epsfig{figure=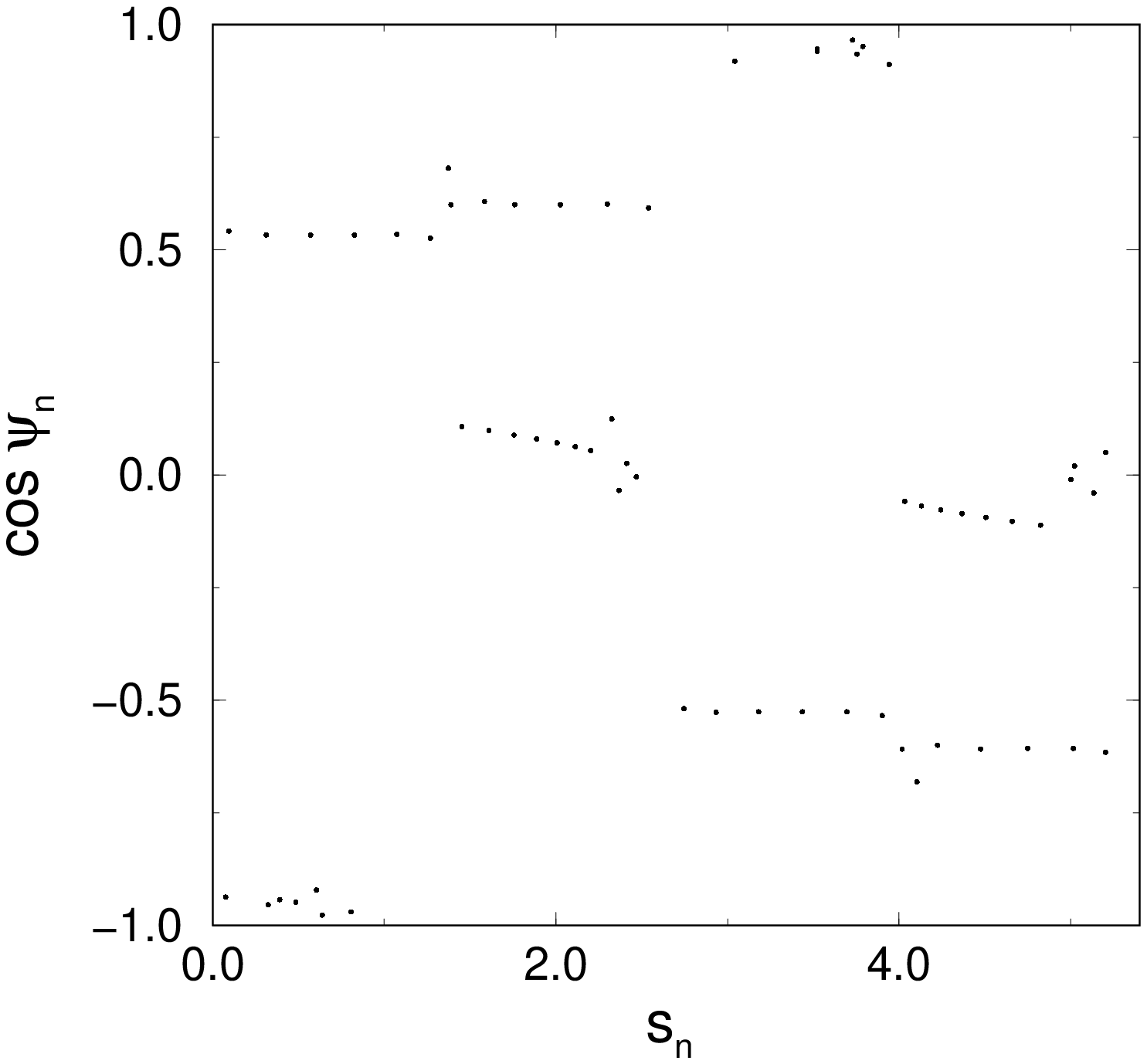,width=7cm}
\caption{\baselineskip=12pt \small
The first 5000 iterates of the bounce map $B$ for $\varepsilon=0.01$
(left panel) starting from a random initial condition, and the 
last 5000 out of a trajectory of $10^7$ collisions (right panel).
}
\label{bmap}
\end{figure}

As remarked in the Introduction, the asymptotic behavior of the system
is trivial, namely all trajectories eventually approach a periodic
orbit. However, before reaching the asymptotic regime, the system 
exhibits a long transient, up to some number of collisions $n_c$ (see later 
for an appropriate definition of $n_c$) during which the
dynamics looks nontrivial, as if the system were
chaotic. The two regimes are illustrated in Fig.\ \ref{bmap}, for
case $\veps=0.01$, which has $n_c\simeq 3\times
10^4$. The left and right panel of the figure report, respectively, the
first and the last 5,000 iterates of the map, out of a trajectory of
$10^7$ collisions. Clearly, during the transient, 5,000 iterates
are sufficient to roughly cover the square, while asymptotically one 
is left with few isolated points.
A closer inspection, by means of hystograms of the density of points in
the square (not reported here) shows that, for smaller $\veps$ and 
correspondingly larger $n_c$, the iterates of the map during the transient
fill more and more uniformly the square.

\begin{figure}
\centering\epsfig{figure=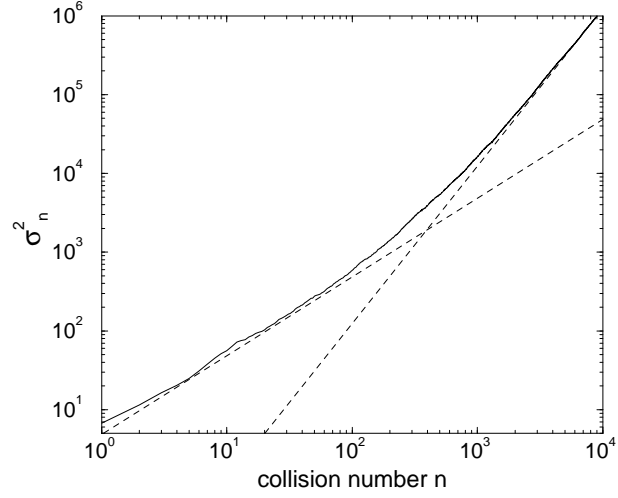,width=8cm}
\caption{\baselineskip=12pt \small
The variance $\sigma_n^2$ for $\varepsilon=0.075$ and ensemble size $N=300$
shows a crossover from diffusive to ballistic behaviour at $n\approx n_c$}
\label{crossover}
\end{figure}

\begin{figure}
\centering\epsfig{figure=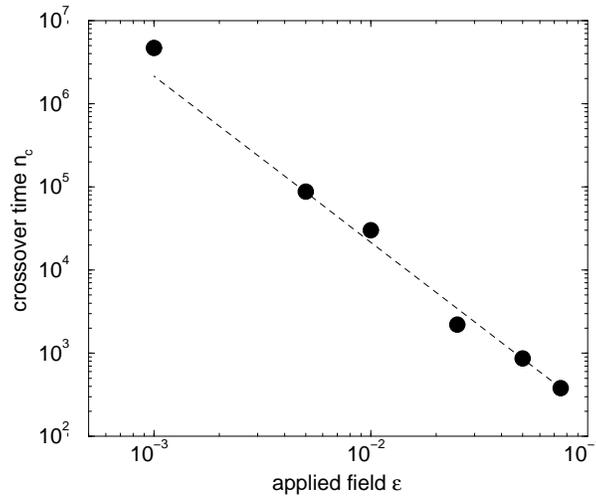,width=8cm}
\caption{\baselineskip=12pt \small
The crossover time $n_c$ as a function of $\varepsilon$.
The dashed line corresponds to $n_c = 2.15~ \varepsilon^{-2}$.}
\label{translen}
\end{figure}

A more quantitative characterization of the two dynamical regimes is 
achieved by considering a large set of initial data $\{ (x_0^{(i)}, 
y_0^{(i)}, p_{x0}^{(i)}, p_{y0}^{(i)}) \}_{i=1}^{N}$, picked up at 
random with uniform distribution in the phase space, and measuring the 
variance $\sigma_n^2=\langle (x_n-x_0)^2+ (y_n-y_0)^2 \rangle$,
where $(x_n,y_n)$ is the actual position of the particle at the $n-$th
collision, and $\langle . \rangle$ denotes averaging over initial data.
This corresponds to an ensemble average, for a non-interacting gas 
of indipendentely thermostatted particles. As is clear from 
Fig.\ \ref{crossover}, one finds a crossover, at $n \approx n_c$, from 
diffusive ($\sigma_n^2 \sim n$) to ballistic ($\sigma_n^2 \sim n^2$)
behaviour, the latter corresponding to the approaching asymptotic
periodic orbit. This is in fact our definition of $n_c$. 
By varying $\veps$, one finds that $n_c$ is  inversely proportional to 
$\veps^2$, see Fig.\ \ref{translen}.
The divergence of the transient for $\veps \rightarrow 0$ is fully
consistent with the conclusions of \cite{CPD}, according to which genuine
diffusive behaviour is found for the wind-tree model without an applied field. 
As already mentioned in the Introduction, this seems to contradict, or 
at least to 
weaken, the claim (see \cite{GBetc}) that observing diffusive behaviour 
in a given physical system is a good indicator that the corresponding 
molecular dynamics is chaotic.

Similarly to the case of the Lorentz gas, the presence of the field induces 
also here an average drift of the particle along the field direction, so we 
can define the ``current" in the system as
\be
\langle p_x \rangle_t = \frac{1}{Nt} \sum_{i=1}^N 
\int_0^t p_x^{(i)}(s) ~d s ~,
\label{avepx}
\ee
i.e.\ as the ensemble average of the time average 
of the component of the particle momentum along the field direction.
All the cases we considered displayed a well-defined positive current 
$\langle p_x \rangle_t > 0$, both for moderately 
long $t$ (in the quasi-stationary transient state) and for very 
long $t$ (in the asymptotic state).

\section{The fluctuation relation}

The discussion of the previous section can be summarized by saying that,
for generic initial conditions, the particle spends a considerably long
time exploring practically all the available phase space, until it
eventually reaches the periodic orbit. Moreover, the time duration of
the transient $T_c=\tau_0 n_c$, $\tau_0$ being the mean flight time,
diverges as $\veps^{-2}$ upon decreasing the field. So, if we restrict
the attention to time scales shorter than $T_c$, it makes sense to study
the statistical properties of the fluctuations of a given observable,
and compare the result with the prediction of the GCFT. To this purpose,
we first need to adapt the latter to the present case. Let
$T > 0$, $M \in \zN$ and take $\zt = T/M$. In our framework, the GCFT
may be replaced by the following conjecture, based on our numerical
observations:

\vskip 1truemm \noindent
{\it {\bf Conjecture:} Consider a periodic billiard with flat scatterers,
which is ergodic at equilibrium. Let the particles be subject to 
external driving and to a Gaussian thermostat. Then, there is a critical
time $T_c$ such that for $T \ll T_c$,
and for sufficiently large $M$ and $\zt$, the following holds:
\be
{1\over \tau \langle p_x\rangle_{_T}}\,
\ln {\pi_\tau (z) \over \pi_\tau (-z)} \;=\; \varepsilon z
\,+\, o(\varepsilon z) \quad ,
\label{flufo}
\ee
where $\pi_\tau$ denotes the probability distribution of the quantity
\be
z \;=\;{\langle p_x\rangle_\tau \over \langle p_x\rangle_T }\quad ,
\ee
computed by subdividing a simulation of length $T$
in segments of length $\zt$, and by recording the 
observed frequency of occurrence of the values $z$.} 
\vskip 1truemm

In the original formulation of Ref.\ \cite{GC}, the fluctuation relation
holds in the limit of large $T$ and $\tau$, for fixed $\veps$. Instead
equation (\ref{flufo}) makes sense only for $T$ and $\tau$ finite.
Alternatively, the limits of large times and of small $\veps$ should be taken
simultaneously, with $T$ well within the transient $T_c\sim\veps^{-2}$, 
when the observables are actually fluctuating.
The correction
term $o(\varepsilon z)$ accounts for the observation that reducing
$\varepsilon$ at fixed $z$, or reducing $z$ at fixed $\varepsilon$,
the left hand side of eq.\ (\ref{flufo}) is well approximated by
$\varepsilon z$ itself.

Under the above limitations, we checked the validity of Eq.(\ref{flufo}) 
for several values of $\veps$. In the numerical computations, it would be 
convenient to consider small fields as they correspond to larger $T_c$
and entail better statistics for $z$ on each run. 
On the other hand, the accuracy of the simulations worsens with decreasing 
field, as usual in nonequilbrium molecular dynamics (see, e.g., \cite{CR}), 
thus a compromise between these contrasting needs is required.
The chosen $\veps$ values range between $10^{-3}$ and $10^{-2}$ and are given 
in the figures along with the values of $\tau$. Incidentally, note that 
these values suffice to go well beyond the linear regime of irreversible 
thermodynamics, which for models of this class can be estimated to correspond 
to fields of $10^{-6}$ or smaller (cf. Ref.\cite{CR}, point 2 of Discussion).

As usually done for billiards \cite{BGG}, 
we decided to cut our long trajectories in segments of a fixed number $n_\tau$
of collisions, rather than of a given duration. Hence, the values of $\zt$ 
reported in the figures are to be understood as the average times necessary 
to undergo $n_\tau$ subsequent collisions with the scatterers. At 
variance with \cite{BGG}, but similarly to \cite{BCL,RS}, we did not 
decorrelate the successive trajectory segments in order to have better 
statistics. A further average over an ensemble of $N\sim 10^5$ 
independent trajectories was also performed. 
 
\begin{figure}
\centering\epsfig{figure=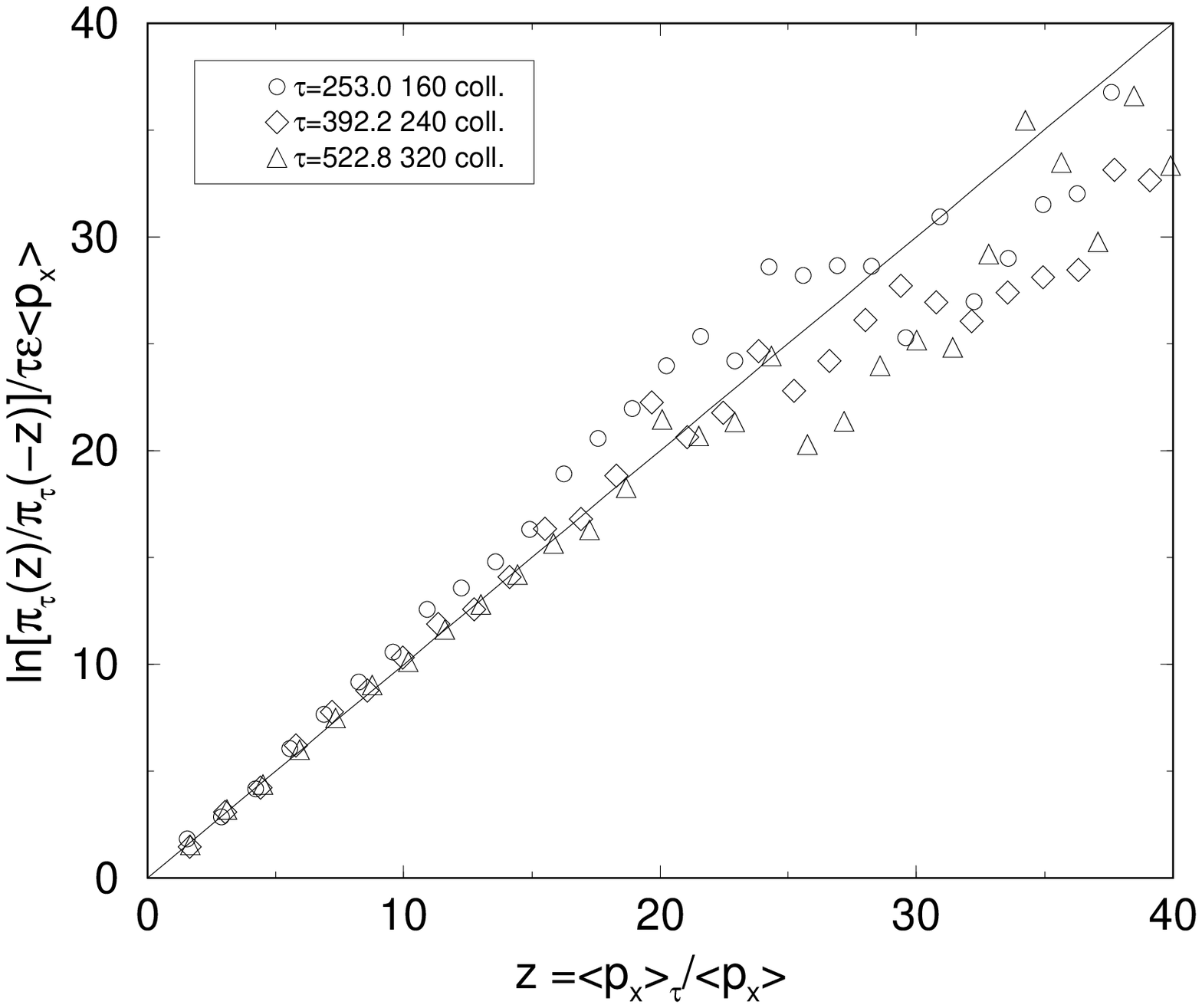,width=7cm}
\centering\epsfig{figure=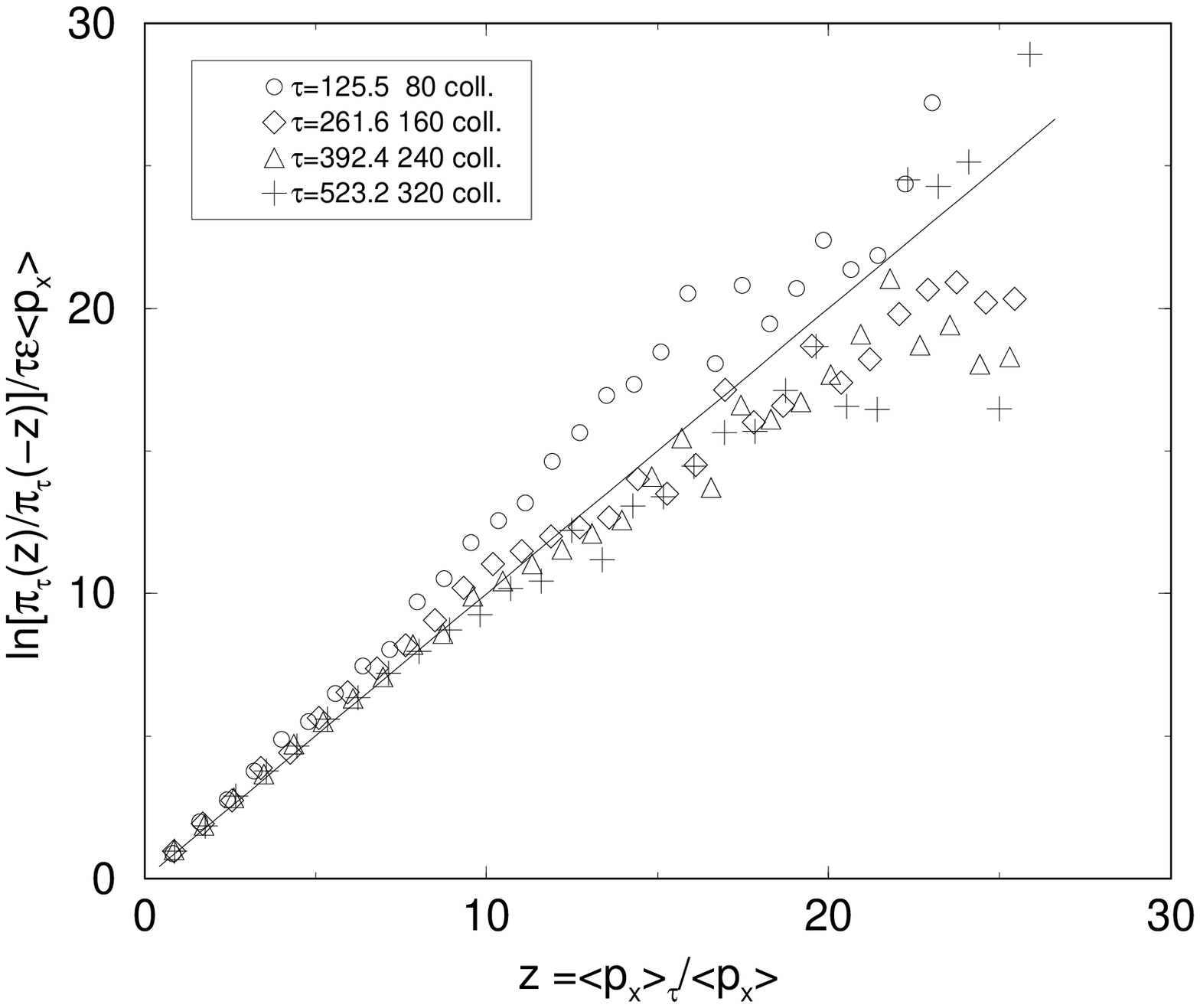,width=7cm}
\caption{\baselineskip=12pt \small
Test of the fluctuation relation for $\varepsilon=0.005$ (left) and
$\varepsilon=0.01$ (right): the solid line is the theoretical prediction
(\protect\ref{flufo}). Large $z$-values are more affected by statistical
errors but its range of validity of it is larger for smaller field.
}
\label{verifica}
\end{figure}

\begin{figure}
\centering\epsfig{figure=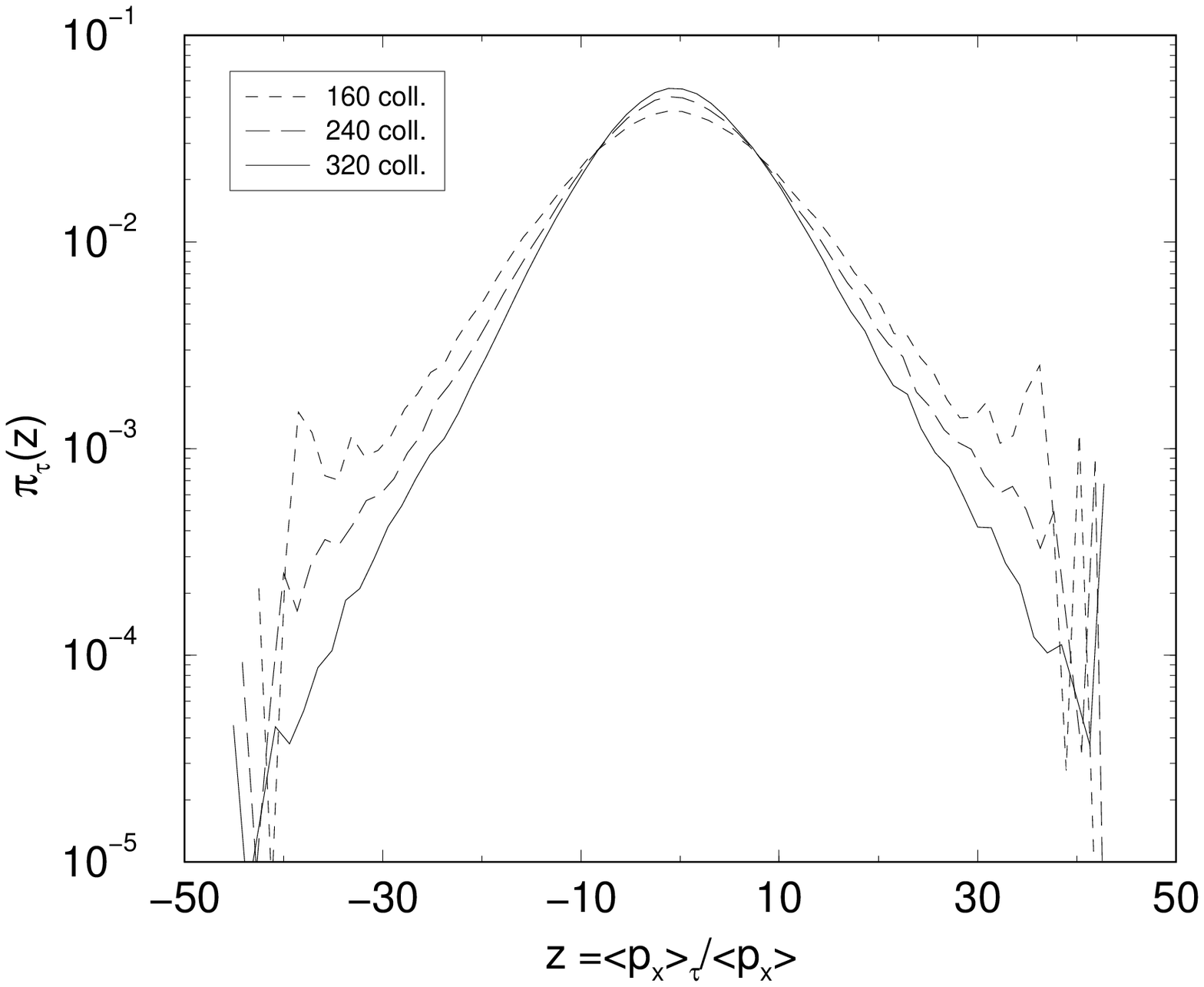,width=7cm}
\centering\epsfig{figure=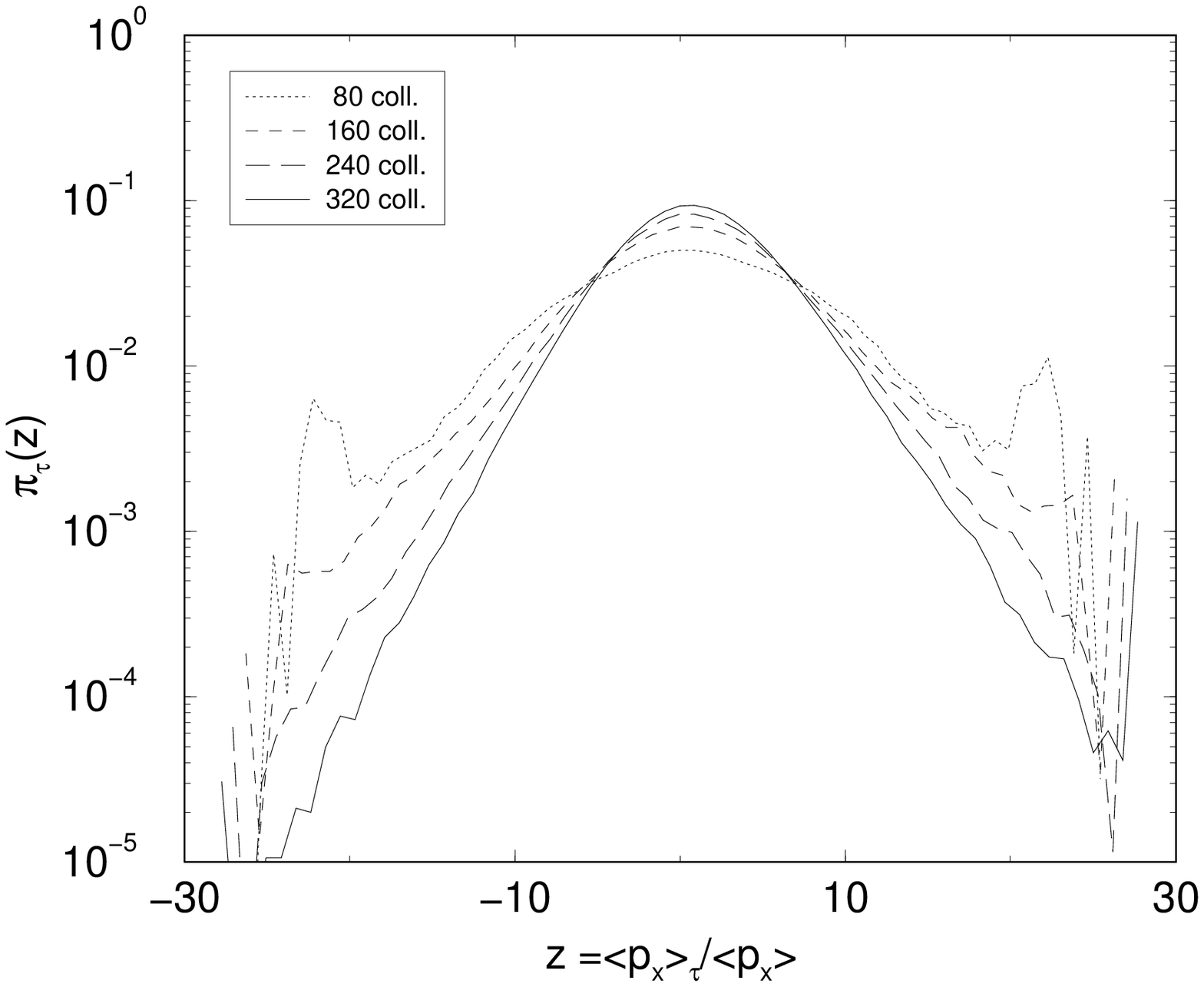,width=7cm}
\caption{\baselineskip=12pt \small
The distribution $\pi_\tau(z)$ of the fluctuation of the current for 
$\varepsilon=0.005$ (left) and $\varepsilon=0.01$ (right).}
\label{distribution}
\end{figure}

Some typical results, reported in Fig.\ \ref{verifica}, seem to
vindicate the above conjecture. In particular, the agreement 
with eq.(\ref{flufo}) improves with increasing $\tau$, 
and the range of validity of the formula gets wider  
by diminishing the field. Fig. \ref{distribution} illustrates the 
non-gaussian nature of the probability distribution for the same 
values of the parameters. Indeed, $\pi_\tau (z)$ appears to be slightly 
asymmetric and to have approximately exponential tails. 

The conclusion we draw is that, in spite of the absence of any source of 
chaoticity, the complex (diffusive) dynamics of our system in 
the transient states is almost indistinguishable from genuine chaotic 
dynamics. 
One may therefore wonder about the origin of the randomness in the wind-tree. 
A possible explanation is the following: it frequently happens
that out of two nearby trajectories, one collides with a scatterer,
while the other just passes by, without being scattered. 
This results in a separation of orbits for short times until, 
eventually, dissipation takes over and both orbits converge to the 
asymptotic one.  
This effect can be easily probed by computing the finite-time Lyapunov 
exponent $\lambda_3(n)$, for $n \ll n_c$. The latter measures precisely the
average separation of close enough initial points after a fixed number of 
collisions. The results reported in  Table 1 show that the time scales over 
which the finite-time exponent appears to be positive covers all the time
range of validity of the fluctuation relation (cf.\ Fig.\ \ref{verifica}). 

\begin{table}
\begin{center}
\begin{tabular}{||r|r|r||} \hline
$n $&$ \veps=0.01 $&$ \veps=0.005 $\\
\hline
$4    $&$ 3.62 ~10^{-1}  $&$ 3.60 ~10^{-1} $\\
\hline
$9    $&$ 2.13 ~10^{-1}  $&$ 2.14 ~10^{-1} $ \\
\hline
$19   $&$ 1.19 ~10^{-1}  $&$ 1.22 ~10^{-1} $ \\
\hline
$39   $&$ 6.96 ~10^{-2}  $&$ 6.87 ~10^{-2} $ \\
\hline
$79   $&$ 4.26 ~10^{-2}  $&$ 3.96 ~10^{-2} $ \\
\hline 
$159  $&$ 2.99 ~10^{-2} $&$ 2.39 ~10^{-2} $ \\
\hline
$319  $&$ 2.44 ~10^{-2} $&$ 1.67 ~10^{-2} $ \\
\hline
$639 $&$ 2.01 ~10^{-2} $&$ 1.40 ~10^{-2} $ \\
\hline
\end{tabular}
\caption{Finite time Lyapunov exponent $\zl_3(n)$, for different  
number of collisions $n$. The ensemble size is $N=10^5$. 
The trend for larger ensembles is similar. We observe that only for
$n \gg n_c$, $\zl_3(n)$ will change sign.}
\end{center}
\end{table}

\section{Concluding remarks}

Our analysis indicates that the consequences of the CH are valid 
(albeit in a restricted sense) for non-chaotic and reversible 
particle systems, provided that their trajectories are sufficiently 
``unpredictable'' for relatively long times. In this respect, the difference
between real chaotic systems and our wind-tree seems only to consist of 
the possibility to extend to infinite times the validity of the CH and of 
its consequences. Similar limitations 
on observation times are usual in statistical mechanics.

Admittedly, the example discussed here is not completely
generic at least for two reasons. First, it is fairly artificial to start
with a system that is ergodic at equilibrium without being actually 
chaotic (even in a weak sense). Second, although other models similar
to ours can be conceived, in general one expects the asymptotic 
state to be chaotic, even in the presence of external fields. 
Nevertheless, our model
is important, in our opinion, 
from a theoretical point of view, because it provides a 
limiting case in which the consequences of the CH can still be 
applied. Accordingly, this also suggests that the CH can be 
successfully applied to systems with slow decay of correlations or long-time 
tails, strongly deviating from the true Anosov systems.
 
As a final remark, we observe that our results are perhaps of some interest 
in the present debate on the relation between the GCFT, and one identity 
previously obtained by Evans and Searles (ESI) \cite{ESI,CohGal99}.
The ESI concerns any time reversible dynamical systems, like ours,
and the Liouville mesaure $\mu_L$ on the phase space $\Omega$
of such systems.  
In particular, let $E_p \subset \Omega$ be the subset of initial 
conditions of trajectories along which the phase space contraction is 
$e^{-p \langle \alpha \rangle T}$ after a time $T>0$,
where $ \langle . \rangle$ represents a (stationary state) average. Then, 
the ESI can be expressed as \cite{CohGal99}:
\be
\frac{\mu_L(E_p)}{\mu_L(E_{-p})} = e^{p \langle \alpha \rangle T} ~.
\label{ESform}
\ee
This equation is formally similar to the fluctuation relation of the GCFT,
especially if one takes very large $T$, so that the computed quantities
characterize the stationary state of the system. However, the GCFT and the 
ESI cannot possibly refer to the same physical quantities. Indeed, if
$T$ is large in the modified Ehrenfest gas, there are no fluctuations of 
the phase space contraction at all. Accordingly, as observed above,
the fluctuation relation does not apply. On the
contrary, the ESI (\ref{ESform}) retains its meaning, showing that
the quantity $\mu_L$ on the left hand side of the
equation cannot be interpreted as probability of flucutations.
The ESI, instead, correctly gives a relation for the 
probabilities of ``trajectory hystories'' with opposite phase 
space contractions. 

\section*{Acknowledgements}
Thanks are in order to E.G.D.\ Cohen, C.P.\ Dettmann, G.\ Gallavotti and
H.\ van Beijeren for very useful remarks. This work has been 
supported by GNFM-CNR, by INFM and by MURST (Italy). 

\section*{Appendix: Lyapunov exponents}

In this appendix we give some expressions for the (finite time) 
Lyapunov exponents of the thermostatted Ehrenfest gas.
Let us consider the coordinates $x$, $y$, and $\theta$, where the latter 
is the angle formed by the momentum vector $(p_x,p_y)$ with the $x$ axis. 
Because the kinetic energy of our system is constant, $K=1/2$, the coordinates
$(x,y,\theta)$ suffice to describe the dynamics which, going from 
one collision to the next, can be split in two stages: 
\be
\pmatrix{\theta_0\cr x_0\cr y_0\cr} \quad 
\stackrel{F}{\longrightarrow}\quad 
\pmatrix{\theta_0'\cr x_0'\cr y_0'\cr} \quad 
\stackrel{C}{\longrightarrow}\quad 
\pmatrix{\theta_1\cr x_1\cr y_1\cr} ~.
\ee
Here $F$, the free flight between two successive obstacles, is explicitly
given by (see eqs. (3) and (4) in \cite{lamberto})
\bey
&&\tan{\theta_0'\over 2} = \tan{\theta_0\over 2} \, e^{-\varepsilon 
\tau(\theta_0,x_0,y_0)}
\label{teta}\\
&&x_0' = x_0 - {1\over \varepsilon} \ln {\sin \theta_0' \over \sin \theta_0}\\
&&y_0' = y_0 - {\theta_0' - \theta_0 \over \varepsilon}
\eey
where $\zt$, the flight time, depends on the initial point
$(\theta_0,x_0,y_0)$. In turn, the map $C$ represents the collision with
the scatterer, and is given by
\bey
&&\theta_1 = f(\theta_0'(\theta_0,x_0,y_0))= -\theta_0' \pm 2\beta\\
&&x_1 = x_0' \\
&&y_1 = y_0'
\eey
where $\theta_0'$ is the incidence angle, $(x_1,y_1)$ is the
collision point, and $\beta$  is the internal angle of the rombus.
The $\pm$ sign depends on the side on which the bounce occours, hence
$C$ is piecewise linear in $\theta_0'$.

In order to compute the Lyapunov exponents, we need to evaluate the 
jacobian matrix $J=J_CJ_F$, product of a free-flight $J_F$ and of
a collision part $J_C$, where
\be
J_F = \pmatrix { {\partial \theta_0'\over \partial\theta_0} & 
{\partial\theta_0'\over\partial x_0} & {\partial\theta_0'\over\partial y_0}\cr
{\partial x_0'\over \partial\theta_0} &  
{\partial x_0'\over\partial x_0} & {\partial x_0'\over\partial y_0}\cr
{\partial y_0'\over \partial\theta_0} &
{\partial y_0'\over\partial x_0} & {\partial y_0'\over\partial y_0}\cr} ~,
\quad
\mbox{and} \quad
J_C = \pmatrix { -1 & 0 &  0 \cr
                  0 & 1 & 0 \cr
                  0 & 0 & 1 \cr} ~.
\ee
Taking the product of matrices like $J$ at each collision, along an
entire trajectory, the Lyapunov exponents $\zl_2, \zl_3, \zl_4$
of eq.(\ref{estim})
can be computed. The remaining one (corresponding to the conserved
kinetic energy) is zero.

Now, consider the relation between the phase space 
contraction rate and the particle momentum \cite{GGOR,CR}:
\be
\mbox{div} (\dot{\bf p},\dot{\bf q}) = - \za = - \veps p_x ~,
\label{psc}
\ee
which holds at all times.
This relation implies that the average of div$(\dot{\bf p},\dot{\bf q})$
(whose asymptotic limit is the sum of the Lyapunov
exponents) has sign opposite to that of the current and, for both 
the finite time and the asymptotic Lyapunonv exponents, we obtain:
\be
\frac{\Delta x(n)}{t_n} = - \frac{\zl_1(n) + \zl_2(n) + \zl_3(n)
+ \zl_4(n)}{\veps} ~.
\label{asymlya}
\ee
Here $\Delta x(n)$ is the distance travelled in real 
space in the time $t_n$ corresponding to $n$ collisions, and $\zl_i(n)$
is the $i$-th finite time exponent. 
This result is exact, and does not require
any conditon on the value of the field. It follows that, because a positive
current is observed  in the stationary state as well as in the transient,
the sum of the Lyapunonv exponents is always negative. Therefore,
one exponent at least is negative. Moreover, in the infinite time limit,
we know that two
exponents vanish, leaving some uncertainty only on the value of the 
remaining exponent.
Numerically, we found that the asymptotic value of this 
Lyapunov exponent is either negative or very close to zero.


\newpage

\vspace{-0.5cm}

\end{document}